\documentclass[preprint,epsfig,onecolumn,floats, showkeys]{revtex4}
\usepackage{makeidx}
\usepackage{amsmath}
\usepackage{amssymb}
\usepackage{graphicx}
\usepackage{epsfig}

\begin{document}

\title{Quantum Search Algorithm for Set Operation}
\author{Chao-Yang Pang}
\email{cyp_900@hotmail.com}
\email{cypang@sicnu.edu.cn}
\affiliation{Key Software Lab., Sichuan Normal University, Chengdu 610066, China;\\
College of Mathematics and Software Science, Sichuan Normal
University, Chengdu 610066, China}
\author{Cong-Bao Ding}
\affiliation{College of Physics and Electronic Engineering, Sichuan
Normal University,} \affiliation{Chengdu 610066, China}
\author{Ben-Qiong Hu}
\affiliation{College of Information Management, Chengdu University
of Technology, 610059, China}

\begin{abstract}
The operations of data set, such as intersection, union and
complement, are the fundamental calculation in mathematics. It's
very significant that designing fast algorithm for set operation. In
this paper, the quantum algorithm for intersection is presented. And
its running time is $O\left( \sqrt{\left\vert A\right\vert \times
\left\vert B\right\vert \times \left\vert C\right\vert }\right) $
for set operation $C=A\cap B$, while classical computation needs
$O\left( \left\vert A\right\vert \times
\left\vert B\right\vert \right) $ steps of computation in general, where $%
|.| $ denotes the size of set. The presented algorithm is the
combination of Grover's algorithm, classical memory and classical
iterative computation, and the combination method decrease the
complexity of designing quantum algorithm.The method can be used to
design other set operations also.
\end{abstract}

\keywords{ Set operation, General Grover iteration, Grover's
algorithm}
 \maketitle

%%%%%%%%%%%%%%%%%%%%%%%%%%%%%%%%%%%
\section{Introduction}
The operations of data set, such as intersection operation and union
operation, are fundamental calculation in mathematics. The fast
computation of set operation is very important because it's the base
of many sciences and techniques, such as database, image processing,
signal processing. E.g, database search is based on set operation
and the fast computation of set operation is very important for it.

The computation procedure of set operation on electronic computer is
illustrated as below.

Suppose there are two vector sets $A$ and B, \\$A=\left\{ \left(
1,1,1,1,\right) ,\left( 2,2,2,2\right) ,\left( 1,2,3,4\right)
\right\} $ ,\\ $B=\left\{ \left( 3,3,3,3\right) ,\left(
4,4,4,4\right) ,\left( 1,2,3,4\right) \right\} $ , \\ and the
intersection set $C=A\cap B=\left\{ \left( 1,2,3,4\right) \right\}
$.
\\Firstly, all vectors of set $A$ (or $B$) are stored in
electronic memory and each vector seems to be a record of database.
The computation procedure of set operation $A\cap B$ is that, for
every vector in set $A$, computer fully searches all elements in set
$B$ and matches it. Because sorting multi-dimensional vectors is no
useful for the speedup of search in general, all vectors of set are
unsorted. Thus, the method of full search becomes the necessary
choice to calculate intersection set for electronic computer, which
is low efficient when set has huge size.

In addition, the running speed of I/O (Input/Output) equipment of
classical computer is the efficiency bottleneck in term of arbitrary
classical algorithm \cite{CSArchitecture}.\ It's the computation
procedure of classical computer that loading data into registers
\textit{one by one} via
I/O, then executing calculation instructors \textit{one by one }\cite%
{CSArchitecture}. If a set has huge amount of elements, the process
of loading data will waste the time heavily and it's an efficiency
bottleneck. E.g., server computer for the database search is more
expensive than personal computer, and one important reason is that
advanced I/O is used. If the size of set is huge, set operation
faces the bottleneck, and there is no way to overcome it on
classical computation principle.

Therefore, for the sets with huge size, electronic computer can do
nothing for the requirement of fast computation. We need new
computation principle and new algorithm for set operation.

Fortunately, in the last decade, quantum computation is studied and
many surprising computation properties are revealed so that the
research of quantum computation becomes the one of hottest research
topic currently. One of milestone of quantum computation researcher
is Shor's algorithm for factoring an integer number with polynomial
computation steps, which is believed to be classically impossible
\cite{Shor}. And Grover presented another exciting algorithm for
database search in 1996. Only $O(\sqrt{N})$ steps of computation are
cost by Grover's algorithm to find a marked element in an unsorted
database with size $N$, while $O(N)$ steps are needed for classical
computer \cite{Grover}. The possible speedup of quantum computation
is essentially enabled by the feature of quantum parallelism. This
parallelism computation of quantum computer bases on the
superposition
property of states, which is not possible on electronic computers \cite%
{Nielson}. It's the computation procedure of quantum computer that
data is loaded into the superposition of states, the superposition
is operated by special unitary operation, the amplitude of solution
is increased, and solution is measured out with big probability at
last. The simple quantum computers have already been constructed.
For example, Shor's algorithm has been demonstrated by NMR quantum
computer \cite{L.M.K.} and by optical quantum computer \cite{C.Y.Lu}
to factorize the number 15.

As well known, the elements of set are arbitrary data (or random
data), the size of set is very big possibly, and all data are stored
in electronic memory unorderly and temporarily. If we want to study
the question that how to use quantum computation to perform set
operation, there are three works must be considered at least. The
first work is that how to express the information of a set using
quantum state. The second work is that how to load the information
of a set into quantum state from electronic memory. The matching
function between two elements $f_{c}$ is a computation, such as
judging if two vector is equal. And the third work is that how to
embed the computation $f_{c}$ into quantum search algorithm.

For the first work (i.e., how to express the set information using
state), there are two expression methods currently. One of method is
proposed by Latorre that the information of classical data is
encoded in the amplitude of a state. Latorre used his method to
expression image data and detail information is lost \cite{Jose}.
Latorre's method is useful for image compression, but it's not
suitable to express the information of general set because
distortion of information is not permitted in set operation. The
other quantum expression is proposed by Pang that all elements are
regarded as sequence of database record $\left\{
record_{0,}record_{1,...,}record_{N-1}\right\} $ and the entangled state $%
\frac{1}{\sqrt{N}}\left( \overset{N-1}{\underset{i=0}{\sum
}}|i\rangle _{register1}|record_{i}\rangle _{register2}\right) $ is
used to express the information of set
\cite{QIC,QVQ2,QVQ1,QLS,Ding,QDFT,QDCT}. Two registers are entangled
in Pang's method, and it is no distortion theoretically. In
addition, the operation of set is equivalent to operating the
entangled state.

For the second work (i.e., how to load the set information into
state from electronic memory), the conception of Quantum Loading
Scheme (QLS) should be introduced. QLS is the unitary operation that
loading all information of data set into quantum state from
electronic memory. Nielsen and Chuang
point out briefly that future quantum computer should have QLS \cite[%
Section 6.5]{Nielson}. Vittorio Giovannetti, Seth Lloyd and et.al.,
present a simple QLS instance with few qubits \cite{Seth}. Pang
presents a QLS instance using the path\ interference of molecule
\cite{QLS}. Pang's study shows that, for a vector
$\overrightarrow{a}$ $=\left\{ a_{1},a_{1},...,a_{N-1}\right\} $,
there is a unitary operation $U_{QLS}$ \cite{QLS} such that

\begin{equation*}
\begin{array}{c}
U_{QLS}:|0\rangle _{register1}|0\rangle _{register2}|ancilla\rangle
_{register3}\longrightarrow \\
\frac{1}{\sqrt{N}}\left( \overset{N-1}{\underset{i=0}{\sum
}}|i\rangle
_{register1}|a_{i}\rangle _{register2}\right) |ancilla\rangle _{register3}%
\end{array}%
\end{equation*}%
, where classical data $a_{1,}a_{1,}...a_{N-1\text{ }}$are used as
control signals to flip the particles.

\begin{figure}[tbh]
\epsfig{file=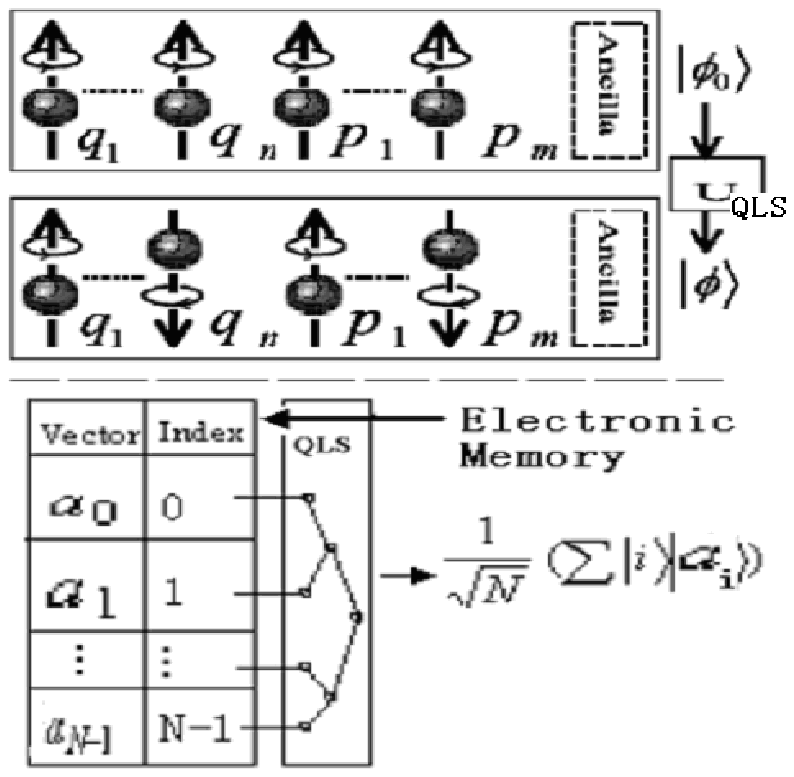,width=10cm,} \caption{The Illustration of
Quantum Loading Scheme (QLS): The function of
QLS is to load all information of vector $\protect\overrightarrow{a}$ $%
=\left\{ a_{1},a_{1},...,a_{N-1}\right\} $ into the superposition of
states
of quantum CPU from electronic memory efficiently. In QLS, classical data $%
a_{1},a_{1},...,a_{N-1}$ are used as control signals to flip the
particles. QLS has time complexity $O(log_{2}N)$ and the I/O
efficiency bottleneck of classical computer is broken by it.}
\label{figQLS}
\end{figure}

Fig.\ref{figQLS} is the illustration of QLS.

The study of Seth Lloyd's group and Pang's study show also that QLS
is fast and has running time $O\left( log_{2}N\right) $, while
classical loading scheme via I/O has running time $O\left( N\right)
$ that is the efficiency bottleneck in term of arbitrary classical
algorithm. Pang also present a variant QLS named unitary operation
$U_{L}$ as below\cite{QLS}

\begin{equation*}
\begin{array}{c}
U_{L}:\frac{1}{\sqrt{N}}\left( \overset{N-1}{\underset{i=0}{\sum
}}|i\rangle
_{register1}|0\rangle _{register2}\right) |ancilla\rangle \longrightarrow \\
\frac{1}{\sqrt{N}}\left( \overset{N-1}{\underset{i=0}{\sum
}}|i\rangle
_{register1}|a_{i}\rangle _{register2}\right) |ancilla\rangle%
\end{array}%
\end{equation*}

The function of operation $U_{L}$ is that loading data from
electronic memory into two entangled registers according indices of
data.

In sum, Nielsen and Chuang points out that QLS has to be existed,
and the study of Seth Lloyd's group and Pang's study both
demonstrate the existence of QLS.

For the third work (i.e., how to embed other computation into
quantum search algorithm), the conception of the general Grover
iteration should be introduced. Additional computation always goes
with the search of database in general. E.g., suppose there is a
database to save the student scores of many subjects. And if we want
to find the student who has maximum average score, the additional
computation $f_{c}$ that calculating average score is also needed.
Famous Grover's algorithm can find a database record according to
the given index, and it the base of many quantum search algorithms.
However, Grover's algorithm is invalid for this kind of search, we
need to improve Grover's algorithm. Pang presents a general Grover
iteration for the
search case with additional computation \cite%
{QIC,QVQ2,QVQ1,QLS,Ding,QDFT,QDCT}, which is derived from the study
of quantum image compression
\cite{QIC,QVQ2,QVQ1,QLS,Ding,QDFT,QDCT}.

The Grover iteration is defined as \cite{Grover,Nielson}
\begin{equation}
G=\left( 2|\xi \rangle \langle \xi |-I\right) O_{f}  \label{eqGI}
\end{equation}%
, where $O_{f}$ is the oracle that flips the phase of state in
Grover
iteration, and $|\xi \rangle =\frac{1}{\sqrt{N}}(\overset{N-1}{\underset{i=0}%
{\sum }}|i\rangle )$.

The \textbf{General Grover Iteration (GGI)} $G_{general\text{ }}$\cite%
{QIC,QVQ2,QVQ1,QLS,Ding,QDFT,QDCT} is defined as

\begin{equation*}
G_{general}=\left( 2|\xi \rangle \langle \xi |-I\right) \left(
U_{L}\right) ^{\dagger }\left( O_{c}\right) ^{\dagger
}O_{f}O_{c}U_{L}
\end{equation*}%
, where $O_{c}$ denotes the other computation oracle for additional
function $f_{c}$.

\begin{figure}[tbh]
\epsfig{file=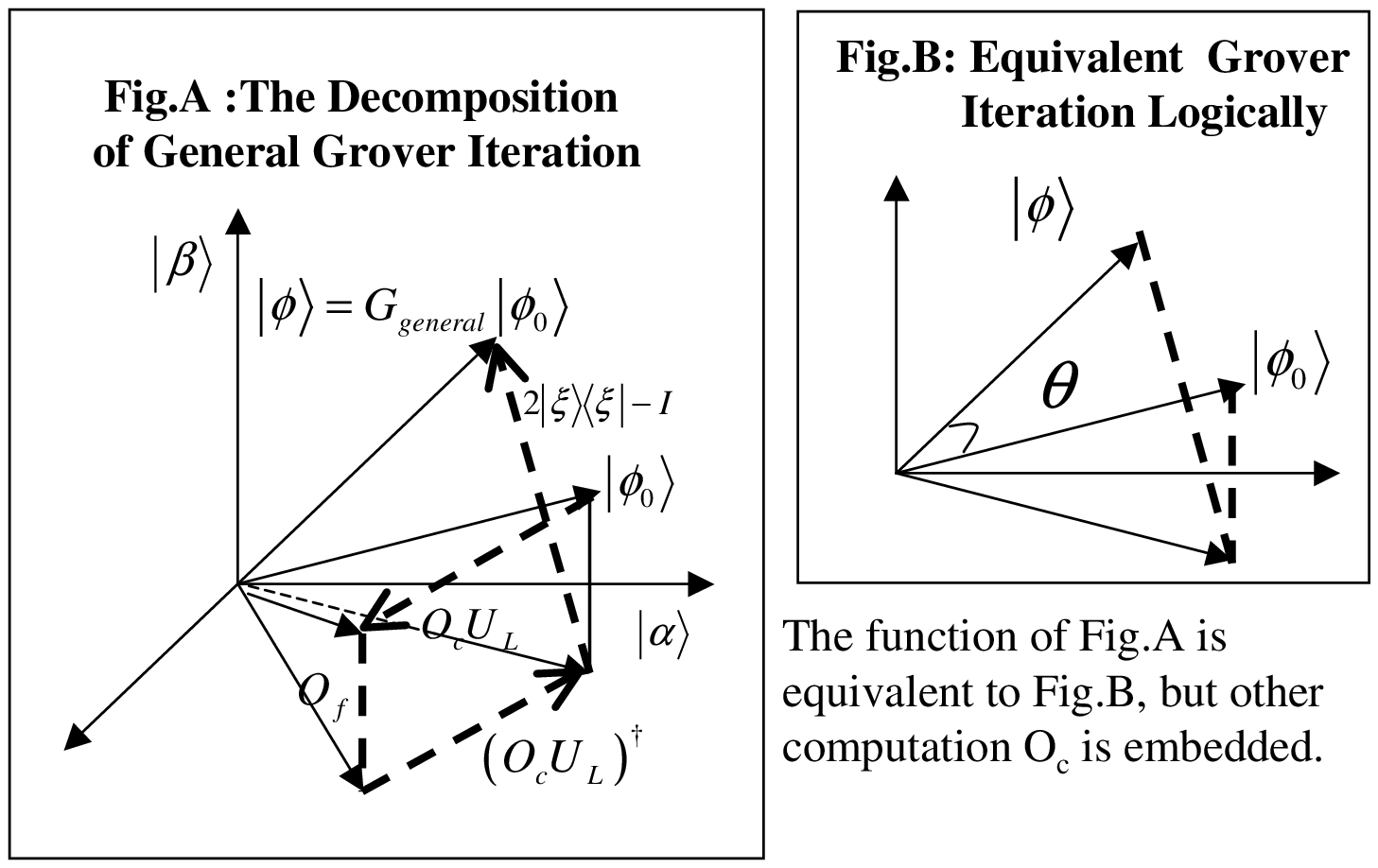,width=10cm,} \caption{The Illustration of
General Grover Iteration} \label{figGGI}
\end{figure}

Fig.\ref{figGGI} illustrates the general Grover iteration.

\bigskip

Similar to Grover iteration, $G_{general}$ act on initial state
$|\xi
\rangle =\frac{1}{\sqrt{N}}\left( \overset{N-1}{\underset{i=0}{\sum }}%
|i\rangle _{register1}|0\rangle _{register2}\right) $ $O\left( \sqrt{N}%
\right) $ times and the solution will be found if the solution is
unique \cite{QIC,QVQ2,QVQ1,QLS,Ding,QDFT,QDCT}.

Grover's algorithm is very useful, and many improved algorithms and
many properties are studied by many experts
\cite{BBHT,Hoye,Long,Biham1,Biham2}. Boyer, Brassard, Hoyer, and Tap
present an improved algorithm named BBHT algorithm in this paper.

Suppose there are sequence of data $T[i]$ ($0\leq i<N$). The various
steps of the BBHT \cite{BBHT} are:

\textbf{Step1.} Initialize $\Gamma =1$ and $\lambda =6/5$ (Any value of $%
\lambda $ strictly between $1$ and $4/3$ would do.)

\textbf{Step2.} Choose $j$ uniformly at random among the nonnegative
integers not bigger than $\Gamma $.

\textbf{Step3.} Apply $j$ iterations of Grover's algorithm starting
from the
state $|\Psi _{0}\rangle =\frac{1}{\sqrt{N}}\overset{N-1}{\underset{i=0}{%
\sum }}|i\rangle $.

\textbf{Step4}. Observe register: let $i_{0}$ be the outcome.

\textbf{Step5.} If $T[i_{0}]=x$, the problem is solved, where $T[i]$
is the sequence of data. And \textbf{exit.}

\textbf{Step6.} Otherwise, set $\Gamma $\ to $min\{\lambda \Gamma ,\sqrt{N}%
\} $ and go to \textbf{step 2.}

The above BBHT algorithm is used to solve the case that the number
of
solutions $t$ is unknown. BBHT algorithm requires that $1\leq t\leq \frac{3}{%
4}N$. If $t>\frac{3}{4}N$, we can applied classical full search
method to got parts of solution efficiently, and call BBHT algorithm
again. The case of no solution is handled by BBHT algorithm also.

BBHT algorithm has time complexity $O\left(
\sqrt{\frac{N}{t}}\right) $. And the probability of finding a
solution is bigger than $\frac{1}{2}$ (i.e., after repeating BBHT
twice or more, a solution will be found with probability $100\%$
approximately).

BBHT algorithm is the combination between quantum algorithm and
classical iteration. And the benefit of this combination lies on
many circuit are saved, while pure quantum algorithm will cost
exponential numbers of circuit when the number of solution is
unknown. BBHT algorithm will be used in this paper \cite{Nielson}.

\section{The Quantum Algorithm for Intersection Operation}

\subsection{Unitary Operation and Data Structure}

Without losing generality, suppose that set is comprised by many
vectors (or
records) and let $A=\left\{ \overrightarrow{a_{0}},\overrightarrow{a_{1}}...%
\overrightarrow{a_{N-1}}\right\} $, where $N=2^{n}$ (otherwise, add
special vector such that $N=2^{n}$). As the same, we have set
$B=\left\{
\overrightarrow{b_{0}},\overrightarrow{b_{1}}...\overrightarrow{b_{M-1}}%
\right\} $, $M=2^{m}$.

The match function between two vectors is defines as
\begin{equation*}
f_{c}\left( \overrightarrow{a_{i}},\overrightarrow{b_{j}}\right)
=\left\{
\begin{tabular}{ccc}
$1$ & $if$ & $\overrightarrow{a_{i}}=\overrightarrow{b_{j}}$ \\
$0$ & \multicolumn{2}{c}{$otherwise$}%
\end{tabular}%
\right.
\end{equation*}
\

The model of intersection operation $C=A\cap B$ is to find two records $%
\overrightarrow{a_{i_{0}}}$ and $\overrightarrow{b_{j_{0}}}$ such that $%
\overrightarrow{a_{i_{0}}}=\overrightarrow{b_{j_{0}}}$ (i.e., $f_{c}(%
\overrightarrow{a_{i}},\overrightarrow{b_{j}})=1$). we have the
following data structure and unitary operation for this model.

\textbf{DS1.} Save set $A$ in electronic memory as a database, and
each vector $\overrightarrow{a_{i}}$ is a record with unique index
$i$. As the same, save set $B$, and each vector
$\overrightarrow{b_{j}}$ has index $j$.

\textbf{DS2.} Construct five registers that have format

\begin{equation*}
|i\rangle _{register1}|j\rangle
_{register2}|\overrightarrow{a_{i}}\rangle
_{register3}|\overrightarrow{b_{j}}\rangle _{register4}|f_{c}(%
\overrightarrow{a_{i}},\overrightarrow{b_{j}})\rangle _{register5}
\end{equation*}

That is, the $1st$, $2nd$, $3rd$, $4th$, $5th$ register is used
respectively
to save index $i$, index $j$, vector $\overrightarrow{a_{i}}$, vector $%
\overrightarrow{b_{j}}$ and the value of match function $f_{c}$.

\textbf{DS3.} Initialize the five registers as zero value:

\begin{equation*}
|0\rangle _{register1}|0\rangle _{register2}|0\rangle
_{register3}|0\rangle _{register4}|0\rangle _{register5}
\end{equation*}

\textbf{DS4.} Construct Hadamard transform:

\begin{equation*}
H:|0\rangle |0\rangle |0\rangle |0\rangle |0\rangle \longrightarrow \frac{1}{%
\sqrt{MN}}\left( \overset{N-1}{\underset{i=0}{\sum }}\underset{j=0}{\overset{%
M-1}{\sum }}\underbrace{|i\rangle |j\rangle }|0\rangle |0\rangle
|0\rangle \right)
\end{equation*}%
,where each ket denotes a register, not a single qubit.

\textbf{DS5.} Construct Quantum Loading Scheme :

\begin{equation}
U_{L}:\frac{1}{\sqrt{MN}}\left( \overset{N-1}{\underset{i=0}{\sum }}\underset%
{j=0}{\overset{M-1}{\sum }}\underbrace{|i\rangle |j\rangle
}|0\rangle |0\rangle |0\rangle \right) \longrightarrow
\frac{1}{\sqrt{MN}}\left(
\overset{N-1}{\underset{i=0}{\sum }}\underset{j=0}{\overset{M-1}{\sum }}%
\underbrace{|i\rangle |j\rangle }|\overrightarrow{a_{i}}\rangle |%
\overrightarrow{b_{j}}\rangle |0\rangle \right)  \label{eqUL}
\end{equation}

The function of $U_{L\text{ }}$is to load data into entangled state
from electronic memory according to index.

\textbf{DS6.} Design oracle $O_{c}$ to compute the value of match function $%
f_{c}$:

\begin{equation*}
O_{c}:\frac{1}{\sqrt{MN}}\left( \overset{N-1}{\underset{i=0}{\sum }}\underset%
{j=0}{\overset{M-1}{\sum }}\underbrace{|i\rangle |j\rangle }|\overrightarrow{%
a_{i}}\rangle |\overrightarrow{b_{j}}\rangle |0\rangle \right)
\longrightarrow \frac{1}{\sqrt{MN}}\left( \overset{N-1}{\underset{i=0}{\sum }%
}\underset{j=0}{\overset{M-1}{\sum }}\underbrace{|i\rangle |j\rangle }|%
\overrightarrow{a_{i}}\rangle |\overrightarrow{b_{j}}\rangle
|f_{c}\left( \overrightarrow{a_{i}},\overrightarrow{b_{j}}\right)
\rangle \right)
\end{equation*}

,where
\begin{equation*}
f_{c}\left( \overrightarrow{a_{i}},\overrightarrow{b_{j}}\right)
=\left\{
\begin{tabular}{ccc}
$1$ & $if$ & $\overrightarrow{a_{i}}=\overrightarrow{b_{j}}$ \\
$0$ & \multicolumn{2}{c}{$otherwise$}%
\end{tabular}%
\right.
\end{equation*}

\textbf{DS7.} Design oracle $O_{f}$:

$%
\begin{array}{c}
O_{f}:\frac{1}{\sqrt{MN}}\left( \overset{N-1}{\underset{i=0}{\sum }}\underset%
{j=0}{\overset{M-1}{\sum }}\underbrace{|i\rangle |j\rangle }|\overrightarrow{%
a_{i}}\rangle |\overrightarrow{b_{j}}\rangle |f_{c}\left( \overrightarrow{%
a_{i}},\overrightarrow{b_{j}}\right) \rangle _{register5}\right)
\longrightarrow \\
\frac{1}{\sqrt{MN}}\left( \overset{N-1}{\underset{i=0}{\sum }}\underset{j=0}{%
\overset{M-1}{\sum }}\left( -1\right) ^{f\left( register5\right) }%
\underbrace{|i\rangle |j\rangle }|\overrightarrow{a_{i}}\rangle |%
\overrightarrow{b_{j}}\rangle |f_{c}\left( \overrightarrow{a_{i}},%
\overrightarrow{b_{j}}\right) \rangle _{register5}\right)%
\end{array}%
$

The above oracle $O_{f}$ is the oracle in Grover's algorithm \cite%
{Grover,Nielson}, which flips the phase of state.

\textbf{DS8.} Construct General Grover Iteration $G_{general}$:
\begin{equation}
G_{general}=\left( 2|\xi \rangle \langle \xi |-I\right) \left(
U_{L}\right) ^{\dagger }\left( O_{c}\right) ^{\dagger
}O_{f}O_{c}U_{L}  \label{eqGGI}
\end{equation}

  The function of operation $G_{general}$ is equivalent to the Grover
iteration (see Fig.\ref{figGGI}).

  If set $C$ has unique element (i.e., $\left\vert C\right\vert =1$), $%
G_{general}$ acting on\\
 $|\Psi _{0}\rangle =\frac{1}{\sqrt{MN}}\left( \overset%
{N-1}{\underset{i=0}{\sum }}\underset{j=0}{\overset{M-1}{\sum }}\underbrace{%
|i\rangle |j\rangle }|0\rangle |0\rangle |0\rangle \right) $ $O\left( \sqrt{%
MN}\right) $ times will generate intersection set. However, the case $%
\left\vert C\right\vert >1$ often happens, where $\left\vert \cdot
\right\vert $denotes the size of set. Therefore, we must design
other improved algorithm to compute all elements of set $C=A\cap B$.

\subsection{\textbf{Subroutine 1: Find An Element in Set} $C=A\cap B$}

\textbf{Step1.} Initial $\Gamma =1$, $\lambda =\frac{6}{5}$.

\textbf{Step2.} Choose $k$ uniformly at random among the nonnegative
integers not bigger than $\Gamma $.

\textbf{Step3.} Apply $k$ times of general Grover iteration
$G_{general}$ starting from the state

$|\Psi _{0}\rangle =\frac{1}{\sqrt{MN}}$ $\left( \overset{N-1}{\underset{i=0}%
{\sum }}\underset{j=0}{\overset{M-1}{\sum }}\underbrace{|i\rangle |j\rangle }%
|0\rangle |0\rangle |0\rangle \right) |ancilla\rangle $.

\textbf{Step4. }Observe the first and second register: let $i_{0}$ and $j_{0%
\text{ }}$be the output.

\textbf{Step5. }If
$\overrightarrow{a_{i_{0}}}=\overrightarrow{b_{j_{0}}}$, return
result $i_{0}$ and $j$, and exit.

\textbf{Step6. }Otherwise, set $\Gamma $ to $min\left\{ \lambda \Gamma ,%
\sqrt{MN}\right\} $ and go to \textbf{step2}.

Subroutine 1 is similar to BBHT algorithm, and the main different
between the two algorithms is that Grover iteration is replaced by
general Grover iteration $G_{general}$.

Similar to BBHT algorithm, we assume that $1\leq |C|\leq
\frac{3}{4}|A|\times
|B|$ in this paper, where $|.|$ denotes the size of set. If $|C|>\frac{3}{4}%
|A|\times |B|$, we can applied classical full search method to got
parts of solutions efficiently, and call subroutine 1 again. Similar
to BBHT, the case $A\cap B=\varnothing $ (i.e., empty set) is
handled by subroutine 1.

Similar to BBHT algorithm, subroutine 1 has time complexity $O\left( \sqrt{%
\frac{\left\vert A|\times |B\right\vert }{\left\vert A\cap B\right\vert }}%
\right) $.

Similar to BBHT algorithm, the output of subroutine 1 is a solution
or not a
solution. And the probability that output is a solution is bigger than $%
\frac{1}{2}$, and the probability that output is not\ a solution is
less than $\frac{1}{2}$. Repeating subroutine 1 twice, a solution
will be obtained.

\subsection{Quantum Search Algorithm for $C=A\cap B$ (Q\_Intersection)}

\textbf{Step1.} $C=\varnothing $ (i.e., empty set) and
$nFlag=FALSE$; Save all elements of set $A$ and $B$ in a database,
each element is a record. And the database is in electronic memory.

\textbf{Step2.} while ($nFlag=FALSE$ )

$\{$

\ \ \ \textbf{Step2-1.} Call subroutine 1 to find a solution $%
\overrightarrow{a_{i_{0}}}=\overrightarrow{b_{j_{0}}}$;

\ \ \ \textbf{Step2-2. }If there is no output from subroutine 1, $nFlag=TRUE$%
;

\ \ \ \ \ \ \ \ \ \ \ If $\overrightarrow{a_{i_{0}}}\notin C$, $%
C\longleftarrow C\cup \left\{ \overrightarrow{a_{i_{0}}}\right\} $.
And
update the database such that the records of vector $\overrightarrow{%
a_{i_{0}}}$ and $\overrightarrow{b_{j_{0}}}$ are different from all
records
and the two vectors are also different. Continue. (Notice: The case $%
\overrightarrow{a_{i_{0}}}\in C$ will not happen in next calling
subroutine 1 because database is updated.)

$\}$

\textbf{Step3.} Call subroutine 1 to find a solution again. If there
is no output from subroutine 1, halt the algorithm. Otherwise, let
$nFlag=FALSE$ and go to \textbf{step2}.

\subsection{\textbf{The Analysis of Time Complexity for Q\_Intersection}}

\textbf{Conclusion:} Algorithm Q\_Intersection has time complexity
$O\left( \sqrt{\left\vert A\right\vert \times \left\vert
B\right\vert \times \left\vert C\right\vert }\right) $, where
$\left\vert \cdot \right\vert $ denotes the size of set.

\textit{Proof: }Similar to BBHT algorithm, the case $\left\vert
C\right\vert =0$ is handled by this algorithm, and running time is
$O(\sqrt{MN})$. The following discussion is under the condition
$\left\vert C\right\vert \geq 1$.

Before firstly calling subroutine 1, there are $t=\left\vert
C\right\vert =\left\vert A\cap B\right\vert $ numbers of unknown
solutions. The sizes of set $A$ and $B$ are both constant during the
whole calculation. Thus, the
scale of problem is $t=\left\vert C\right\vert $. Suppose we need $%
I_{t}=I_{\left\vert C\right\vert }$ steps of computation to obtain
all
solutions. During the first computation of calling subroutine 1, $c\sqrt{%
\frac{MN}{\left\vert C\right\vert }}$ steps of computation are cost, where $%
c $ denotes a constant, $|A|=N$ and $|B|=M$. The output of
subroutine 1 is a solution or not a solution. After executing
step2-1 (i.e., subroutine 1), two cases are happened. And the first
case is that the output of subroutine 1 is a solution, and the
second case is that the output is not a solution. The probability of
the first case $P_{case1}$ is bigger than $\frac{1}{2}$,
while the probability of the second case $P_{case2}$\ is less than $\frac{1}{%
2}$. That is, there is a real number $\varepsilon $ ($0\leq
\varepsilon \leq
\frac{1}{2}$) such that $P_{case1}=\frac{1}{2}+\varepsilon $ and $P_{case2}=%
\frac{1}{2}-\varepsilon $ . When the first case happen, the scale of
problem becomes $t-1=\left\vert C\right\vert -1$, and
$I_{t-1}=I_{\left\vert C\right\vert -1}$ computation steps will be
needed for all of remnant
calculations. When the second case happen, the scale of problem is still $%
\left\vert C\right\vert $, and $I_{t}=I_{\left\vert C\right\vert }$
computation steps will be needed again.

When secondly calling subroutine 1, the situation is same. When the
scale of problem (i.e., the number of unknown solutions) is $t$,
$c\sqrt{\frac{MN}{t}}$ steps of computation are cost for calling
subroutine 1, and the number of remnant steps is
$(\frac{1}{2}+\varepsilon )I_{t-1}+(\frac{1}{2}-\varepsilon )I_{t}$.
That is,
\begin{equation*}
I_{t}=c\sqrt{\frac{MN}{t}}+(\frac{1}{2}+\varepsilon )I_{t-1}+(\frac{1}{2}%
-\varepsilon )I_{t}
\end{equation*}

With the iterative computation increasing, the scale of problem $t$
become smaller. When $t=1$, $O(\sqrt{MN})$ steps of computation will
be cost by subroutine 1, i.e., $I_{1}=c_{1}\sqrt{MN}$, where $c_{1}$
is a constant.

The time complexity can by analyzed by the above way. Therefore, the
following recursion equation is obtained to calculate time complexity: $\ \ $%
\begin{equation}
\left\{
\begin{array}{ccc}
I_{t} & = & c\sqrt{\frac{MN}{t}}+(\frac{1}{2}+\varepsilon )I_{t-1}+(\frac{1}{%
2}-\varepsilon )I_{t} \\
&  & I_{1}=c_{1}\sqrt{MN} \\
&  & 1\leq t\leq \left\vert C\right\vert ,0\leq \varepsilon \leq \frac{1}{2}%
\end{array}%
\right.  \label{eqIterationNumber}
\end{equation}

, where $I_{t\text{ }}$denotes the number of computation steps when
there are $t$ number of unknown solution.

By Eq.\ref{eqIterationNumber}, we have

\begin{equation}
I_{t}-I_{t-1}=2c\frac{\sqrt{MN}}{\sqrt{t}}-2\varepsilon
(I_{t}-I_{t-1}) \label{Eq2}
\end{equation}

Performing \ $\left( I_{\left\vert C\right\vert }-I_{\left\vert
C\right\vert -1}\right) +\left( I_{\left\vert C\right\vert
-1}-I_{\left\vert C\right\vert -2}\right) +...\left(
I_{2}-I_{1}\right) $, we have

\begin{equation*}
I_{\left\vert C\right\vert }-I_{1}=2c\sqrt{MN}(\overset{|C|}{\underset{i=2}{%
\sum }}\frac{1}{\sqrt{i}})-2\varepsilon (I_{\left\vert C\right\vert
}-I_{1})
\end{equation*}

Because $I_{\left\vert C\right\vert }\geq I_{1}$ and $\varepsilon
\geq 0$, we have

\begin{equation*}
I_{\left\vert C\right\vert }-I_{1}\leq 2c\sqrt{MN}(\overset{|C|}{\underset{%
i=2}{\sum }}\frac{1}{\sqrt{i}})
\end{equation*}

That is,

\begin{equation*}
I_{\left\vert C\right\vert }\leq 2c\sqrt{MN}(\overset{|C|}{\underset{i=2}{%
\sum }}\frac{1}{\sqrt{i}})+I_{1}
\end{equation*}

We have

\begin{equation*}
I_{\left\vert C\right\vert }\leq
2c\sqrt{MN}(\underset{1}{\overset{|C|}{\int
}}\frac{1}{\sqrt{x}}dx)+I_{1}
\end{equation*}

\begin{equation*}
I_{\left\vert C\right\vert }\leq 4c\sqrt{\left\vert C\right\vert
MN}+\left( c_{1}-4c\right) \sqrt{MN}
\end{equation*}

i.e.,

\begin{equation*}
I_{\left\vert C\right\vert }\leq
(4c+\frac{c_{1}-4c}{\sqrt{\left\vert C\right\vert
}})\sqrt{\left\vert C\right\vert MN}
\end{equation*}

Because $\left\vert C\right\vert \geq 1$, we have

\begin{equation*}
\left\{
\begin{array}{c}
\begin{array}{ccc}
I_{\left\vert C\right\vert }\leq c_{1}\sqrt{\left\vert C\right\vert
MN} &
\text{if} & c_{1}\geq 4c%
\end{array}
\\
\begin{array}{ccc}
I_{\left\vert C\right\vert }\leq 4c\sqrt{\left\vert C\right\vert MN} & \text{%
if} & c_{1}<4c%
\end{array}%
\end{array}%
\right.
\end{equation*}

Thus, there is a constant $\lambda >0$ such that
\begin{equation*}
I_{\left\vert C\right\vert }\leq \lambda \sqrt{\left\vert
C\right\vert MN}
\end{equation*}

That is, \
\begin{equation}
I_{\left\vert C\right\vert }=O(\sqrt{\left\vert A\right\vert \times
\left\vert B\right\vert \times \left\vert C\right\vert })
\label{eqIC}
\end{equation}

Formula \ref{eqIC} shows that Q\_Intersection algorithm has time complexity \\ $%
O\left( \sqrt{\left\vert A\right\vert \times \left\vert B\right\vert
\times \left\vert C\right\vert }\right) $. That is, only $O\left(
\sqrt{\left\vert
A\right\vert \times \left\vert B\right\vert \times \left\vert C\right\vert }%
\right) $ steps of computation are needed to calculate intersection set $%
C=A\cap B$, while $O\left( \left\vert A\right\vert \times \left\vert
B\right\vert \right) $ steps are needed for classical computation.
The quantum algorithm Q\_Intersection is fast than classical method
by $O\left(
\sqrt{\frac{\left\vert A\right\vert \times \left\vert B\right\vert }{%
\left\vert C\right\vert }}\right) $ factors.

In addition, the probability of BBHT algorithm to find a solution is
bigger than $\frac{1}{2}$. Similar to BBHT algorithm, The successful
probability of subroutine 1 is bigger than $\frac{1}{2}$. That is,
calling subroutine 1 twice can find a solution with 100\%
probability approximately. Step2 and step 3 of Q\_Intersection
algorithm guarantee the successful probability is close to $100\%$
approximately.

Because $A\cup B=I-A\cap B$, the presented algorithm can be used to
calculate union operation also. In sum, using the method of
Q\_Intersection to perform set operation is possible.

\section{\textbf{Conclusion}}

The set operations, such as intersection, union and complement, are
the fundamental calculation in mathematics. Set operation is the
base of many sciences and techniques, such as database, signal
processing and image processing. E.g., database is based on set
operation. Designing fast algorithm for set operation is
significant.

Full search method is the common method of set operation in general
because sorting multi-dimensional data is not very useful to improve
running speed. Full search has time complexity $O\left( \left\vert
A\right\vert \times \left\vert B\right\vert \right) $ for the
intersection $C=A\cap B$, which is very slow still when the size of
set is huge. Electronic computer loads data into register
\textit{one by one} from memory, and the efficiency bottleneck is
formed.

In this paper, the quantum search algorithm for intersection
operation of set (named Q\_Intersection) is presented, which is the
combination of Grover' algorithm, classical memory, classical
iteration. Using the method of Q\_Intersection, the quantum
algorithms for other set operations can be designed also.

The advantages of Q\_Intersection are listed as below.

1. Q\_Intersection has time complexity $O\left( \sqrt{\left\vert
A\right\vert \times \left\vert B\right\vert \times \left\vert C\right\vert }%
\right) $, while classical algorithm has time complexity $O\left(
\left\vert A\right\vert \times \left\vert B\right\vert \right) $,
where $\left\vert \cdot \right\vert $ denotes the size of set.
Q\_Intersection is fast than classical method by $O\left(
\sqrt{\frac{\left\vert A\right\vert \left\vert B\right\vert
}{\left\vert C\right\vert }}\right) $.

2. All information of data can be loaded into quantum state by
$O\left( log_{2}\left\vert A\right\vert \left\vert B\right\vert
\right) $ steps of computation in Q\_Intersection, and all data is
loaded at a same time, while classical computer can only load data
one by one and $O(\left\vert A\right\vert \left\vert B\right\vert )$
steps are needed. And the efficiency bottleneck of electronic
computer is evaded.

3. In step2-2 of Q\_Intersection, the data in the electronic memory
is often updated according to the output of quantum algorithm, which
simplifies the design of quantum algorithm. As well known, data in
the superposition of states can not be updated as a given number and
can not be measured when unitary operation acting on this
superposition. This defect make designing quantum algorithm very
difficult. Q\_Intersection shows that the combination between
quantum algorithm and classical memory is useful to decrease the
complexity of designing quantum algorithm.

\begin{acknowledgments}
The first author thanks his teacher prof. G.-C. Guo because the author's
main study methods are learned from his lab. The first author thanks prof.
Z. F. Han's for he guiding the author up till now. The first author thanks
prof. J. Zhang and prof. Z.-L. Pu for their help.
\end{acknowledgments}

\end{document}